\documentclass[4pt,aps,pre,reprint,superscriptaddress
]{revtex4-2} %
\usepackage[dvipsnames]{xcolor}
\usepackage[utf8]{inputenc}
\usepackage[T1]{fontenc}
\usepackage{amsmath,amssymb,amsthm,bm}

\usepackage{hyperref}

\usepackage{dcolumn}
\usepackage{mathptmx}
\usepackage{amsmath,amssymb,mathtools}
\usepackage{physics}
\usepackage{enumitem}
\usepackage[cal=dutchcal,scr=rsfs,frak=esstix,bb=ams]{mathalfa}

\begin{document}
\title{The Constrained Origin of Canonical and Microcanonical Ensembles in Quantum Theory}
\author{Loris Di Cairano}
\affiliation{Department of Physics and Materials Science, University of Luxembourg, L-1511 Luxembourg City, Luxembourg}
\email{l.di.cairano.92@gmail.com}

\date{\today}

\begin{abstract}
In quantum theory, equilibrium statistical mechanics is usually formulated through the canonical ensemble, whose privileged status is tied to the Euclidean continuation of time evolution. The microcanonical ensemble, by contrast, is commonly introduced as a separate spectral construction. In this work we show that this asymmetry is representational rather than structural. We formulate the system in an extended Hilbert space in which time is promoted to an auxiliary clock degree of freedom and physical states are selected by a reparametrization-invariant constraint operator $\hat C = \hat P_T + \hat H$. The corresponding projector $\delta(\hat C)$ provides a single unified object from which both canonical and microcanonical ensembles emerge as complementary projections in the clock sector. In the clock-time representation, a purely imaginary clock separation yields the Euclidean kernel and the canonical partition function. In the conjugate clock-energy representation, the same projector reduces to the spectral operator $\delta(\hat H-E)$ and hence to the microcanonical density of states. The main consequence is structural: canonical and microcanonical statistics need not be introduced as independent constructions, since both are already encoded in the same constrained quantum dynamics.
\end{abstract}

\maketitle

\section{Introduction}

In quantum mechanics and field theory, the canonical ensemble has long appeared as the natural statistical completion of the dynamical formalism. The reason is not merely thermodynamic but structural. Quantum dynamics is directly encoded in the transition amplitude which admits a path-integral representation of the form~\cite{peskin2018introduction}:
\begin{equation*}
    \langle q_f|e^{-i\hat{H}(t_f-t_i)}|q_i\rangle
    \sim
    \int_{q_i}^{q_f} \mathcal Dq \; e^{iS[t_i,t_f;q]},
\end{equation*}
and provides at once both a formulation of dynamics and a bridge between classical and quantum theory through the action functional \(S\). Once this connection is available, the passage to equilibrium statistical mechanics is immediate: after analytic continuation to imaginary time~\cite{kapusta2007finite}, the evolution operator \(e^{-i\hat H(t_f-t_i)}\) becomes \(e^{-\beta \hat H}\), and the canonical partition function follows by tracing over the Hilbert/Fock space,
\begin{equation*}
    Z(\beta)=\mathrm{Tr}\,e^{-\beta \hat H}.
\end{equation*}
In this way, the canonical ensemble is not only technically accessible, but also appears as the direct statistical image of quantum dynamics itself.

The microcanonical ensemble does not share the same immediate status. In the standard formulation, it is typically introduced through the spectral object:
\begin{equation*}
    \Omega(E)=\mathrm{Tr}\,\delta(\hat H-E),
\end{equation*}
or, equivalently, as a sum over the eigenvalues in the spectrum of the Hamiltonian operator. Unlike the canonical operator \(e^{-\beta\hat H}\), which is directly tied to time evolution and its Euclidean continuation, the microcanonical projector \(\delta(\hat H-E)\) does not arise naturally from the standard path-integral chain connecting action, quantization, and thermodynamics. For this reason, the canonical ensemble has historically come to seem not only more convenient but also more fundamental. Its privileged position in quantum theory is therefore rooted as much in technical accessibility as in physical habit.

However, this centrality is not imposed by some fundamental law in physics. It reflects the existence of a particularly powerful identification between time evolution and thermal weights, not the necessity of canonical statistics in all regimes. There are important domains in which this canonical route is incomplete, singular, or conceptually obstructed. In generally covariant systems, and especially in gravitational theories, the very notion of external time becomes problematic, so that the usual identification between imaginary time and inverse temperature is no longer
available in any straightforward way
\cite{Wheeler1968,DeWitt1967,HenneauxTeitelboim1992,Kuchar1992,Rovelli1991,BrownYork1993}.
In systems with exponentially growing densities of states, such as
Hagedorn-type spectra, the canonical partition function may diverge above a critical scale even though the fixed-energy description remains meaningful
\cite{Hagedorn1965,AtickWitten1988,strominger1983microcanonical}. More broadly, systems with long-range interactions or strong correlations can exhibit inequivalence between canonical and microcanonical descriptions, so that the fixed-energy viewpoint becomes essential rather than optional
\cite{Thirring1970,BarreMukamelRuffo2001,BouchetGuptaMukamel2010,defenu2024ensemble}. These
examples do not invalidate the canonical ensemble, but they do show that its historical centrality should not be mistaken for formal inevitability.

A revealing perspective emerges already at the classical level. The central role of time in standard quantum theory is inherited from the Hamilton variational principle~\cite{lanczos2012variational}, where one seeks trajectories connecting two configurations at fixed initial and final times. From this starting point, the time-evolution operator and its path-integral representation arise naturally, and with them, the direct route to canonical thermodynamics. However, Hamilton's principle is not the only variational formulation available~\cite{arnol2013mathematical,landau1960mechanics}. In the Maupertuis-Jacobi principle~\cite{brown1989jacobi}, time is not fixed at the endpoints; instead, the energy is fixed and enters as a constraint. The contrast between these two formulations already suggests that the apparent asymmetry between fixed-temperature and fixed-energy statistical descriptions may not be fundamental but inherited from a prior choice of variational framework.

The point of the present work is precisely to show that this asymmetry is only formal. The apparent primacy of the canonical ensemble in quantum theory is representational rather than structural. Our aim is to identify the common operatorial structure from which both descriptions follow and show that canonical and microcanonical ensembles arise as two different and equally legitimate projections of a more general reparametrization-invariant dynamics~\cite{johns2011analytical,HenneauxTeitelboim1992}.

The key idea is to formulate the system in an extended Hilbert space,
\begin{equation}
    \mathcal H_{\mathrm{ext}}
    =
    \mathcal H_T\otimes \mathcal H_{\mathrm{QM}},
\end{equation}
where \(\mathcal H_T\) is an auxiliary clock sector. Physical states are then selected by a reparametrization-invariant constraint
\begin{equation}
    \hat C=\hat P_T+\hat H\simeq0,
\end{equation}
and the central object becomes the corresponding projector:
\begin{equation}
    \delta(\hat C)
    =
    \frac{1}{2\pi}\int_{-\infty}^{+\infty} d\alpha\,e^{i\alpha \hat C}.
\end{equation}
Once this constrained kernel is given, the canonical and microcanonical
ensembles are recovered as two complementary representations in the clock
sector. In the clock-time representation, one recovers the Euclidean kernel and hence the canonical ensemble. In the conjugate clock-energy representation, one recovers instead the spectral projector \(\delta(\hat H-E)\) and hence the microcanonical ensemble.

We show that the conceptual content of this construction is already anticipated by 
classical theory. Parametrized Hamiltonian dynamics provides a unified
framework in which fixed-time and fixed-energy descriptions arise as different boundary-value realizations of the same constrained system. The quantum formalism developed below is the operatorial counterpart of this classical unity. What appears in the standard formulation as a hierarchy between canonical and microcanonical ensembles is reinterpreted here as a choice of representation of one and the same underlying kernel.

The message is therefore sharp. Canonical and microcanonical ensembles are not independent constructions built directly at the level of \(\hat H\); they are two projections of the same reparametrization-invariant kernel in an extended Hilbert space. The canonical ensemble appears privileged only because the standard formalism gives direct access to it through time evolution and Euclidean continuation. Once the theory is reformulated at the level of constrained dynamics, that privilege is seen to be representational rather than fundamental.

\section{Parametrized Hamiltonian theory in classical systems: ensembles as boundary selection}
\label{sec:root-action}

Classical Hamiltonian dynamics admits a natural formulation on an extended phase space in which time is treated as a canonical variable~\cite{rovelli2015covariant,johns2011analytical}. This makes reparametrization invariance manifest and provides a common variational origin for both the Hamilton and Maupertuis--Jacobi principles. The main point of this section is that the canonical and microcanonical descriptions already appear, at the classical level, as two realizations of the same constrained dynamics, distinguished by different boundary data.

\subsection{Parametrized action on extended phase space}
\label{subsec:root-param}

Let \((q^i,p_i)\), \(i=1,\dots,D\), be canonical coordinates with the Hamiltonian:
\begin{equation}
    H(p,q)=\sum_{i=1}^D \frac{p_i^2}{2m}+V(q).
    \label{eq:Ham-standard}
\end{equation}
We enlarge the phase space by promoting time \(t\) to a canonical coordinate with conjugate momentum \(\pi_t\). The extended phase space is then coordinatized by~\cite{}:
\begin{equation*}
    (q^i,p_i;t,\pi_t),
\end{equation*}
with a canonical one-form and symplectic form:
\begin{equation}
    \theta_{\rm ext}=p_i\,dq^i+\pi_t\,dt,
    \qquad
    \omega_{\rm ext}=dq^i\wedge dp_i+dt\wedge d\pi_t.
\end{equation}

The dynamics is encoded by the first-class constraint:
\begin{equation}
    \mathscr C(q,p,t,\pi_t):=H(p,q)+\pi_t\approx 0,
    \label{eq:root-constraint}
\end{equation}
which defines the physical hypersurface in the extended phase space. Introducing an arbitrary parameter \(\sigma\) along the trajectory and a Lagrange multiplier \(\tilde N(\sigma)\), one obtains the parametrized action~\cite{hartle1984path,brown1989jacobi}:
\begin{equation}
    S=
    \int_{\sigma_a}^{\sigma_b} d\sigma
    \left[
    p_i\frac{dq^i}{d\sigma}
    +\pi_t\frac{dt}{d\sigma}
    -\tilde N(\sigma)\big(H(p,q)+\pi_t\big)
    \right].
    \label{eq:S-root}
\end{equation}
This action is invariant under reparametrizations
\(\sigma\mapsto \sigma'=f(\sigma)\), provided \(\tilde N\) transforms as a density:
\begin{equation*}
    \tilde N(\sigma)\,d\sigma=\tilde N'(\sigma')\,d\sigma'.
\end{equation*}

A straightforward variation yields:
\begin{equation}
    \delta S
    =
    \int_{\sigma_a}^{\sigma_b} d\sigma\,(\text{bulk})
    +
    \big[p_i\,\delta q^i+\pi_t\,\delta t\big]_{\sigma_a}^{\sigma_b},
    \label{eq:root-boundary}
\end{equation}
together with the bulk equations:
\begin{align}
  \frac{dq^i}{d\sigma} &= \tilde N(\sigma)\,\frac{\partial H}{\partial p_i},
  &
  \frac{dp_i}{d\sigma} &= -\,\tilde N(\sigma)\,\frac{\partial H}{\partial q^i},
  \label{eq:root-eom-qp}
  \\
  \frac{dt}{d\sigma} &= \tilde N(\sigma),
  &
  \frac{d\pi_t}{d\sigma} &= 0,
  \label{eq:root-eom-time}
  \\
  &\qquad H(p,q)+\pi_t =0.
  \label{eq:root-eom-constraint}
\end{align}
Thus, the same parametrized theory contains both the dynamical equations and the energy constraint, while leaving the evolution parameter \(\sigma\) arbitrary.

\subsection{Fixed-time boundary conditions and Hamilton's principle}
\label{subsec:root-Hamilton}

We first consider the variational problem in which both the configuration
variables and time are fixed at the endpoints:
\begin{equation}
    \delta q^i(\sigma_a)=\delta q^i(\sigma_b)=0,
    \qquad
    \delta t(\sigma_a)=\delta t(\sigma_b)=0.
\end{equation}
Under these conditions, the boundary term in \eqref{eq:root-boundary} vanishes, and Eqs.~\eqref{eq:root-eom-qp}--\eqref{eq:root-eom-constraint} follow.

Because the action is reparametrization-invariant, we may choose the gauge:
\begin{equation}
    t(\sigma)=\sigma,
    \qquad
    \sigma\in[t_a,t_b],
\end{equation}
so that
\begin{equation}
    \frac{dt}{d\sigma}=1
    \qquad\Rightarrow\qquad
    \tilde N(\sigma)=1.
\end{equation}
The constraint then provides:
\begin{equation}
    \pi_t=-H(p,q),
\end{equation}
and the parametrized action reduces to:
\begin{equation}
    S_H[p,q;t_a,t_b]
    =
    \int_{t_a}^{t_b} dt\,
    \Big[
    p_i(t)\,\dot q^i(t)-H\big(p(t),q(t)\big)
    \Big],
    \label{eq:SH-recovered-clean}
\end{equation}
which is the standard Hamilton action. The equations of motion become the usual Hamilton equations:
\begin{equation}
  \dot q^i=\frac{\partial H}{\partial p_i},
  \qquad
  \dot p_i=-\frac{\partial H}{\partial q^i}.
\end{equation}
The point is, therefore, structural: the ordinary fixed-time Hamiltonian
description is recovered from the same parametrized constrained theory once one chooses time as gauge and fixes \(q^i\) and \(t\) at the endpoints. At the statistical level, this is the classical setting underlying the canonical description.

\subsection{Fixed-energy boundary conditions and the Maupertuis--Jacobi principle}
\label{subsec:root-MJ}

We now consider a different variational problem, appropriate for fixed energy. In this case, the relevant boundary data are the endpoint configurations and the value of the conserved energy, while the travel time is left free. This is implemented by fixing \(q^i\) and \(\pi_t\) at the endpoints while allowing \(t\) to vary.

To adapt the action to this boundary data, we perform a Routh transformation in the extended phase space and define
\begin{equation}
  \tilde S[q,p,t,\pi_t,\tilde N]
  :=
  S[q,p,t,\pi_t,\tilde N]
  -
  \big[\pi_t(\sigma)\,t(\sigma)\big]_{\sigma_a}^{\sigma_b}.
  \label{eq:Routh-def}
\end{equation}
Using \(\pi_t t'=(\pi_t t)'-\pi_t' t\), this becomes
\begin{equation}
  \tilde S
  =
  \int_{\sigma_a}^{\sigma_b} d\sigma\,
  \left[
    p_i\frac{dq^i}{d\sigma}
    -t\,\frac{d\pi_t}{d\sigma}
    -\tilde N(\sigma)\big(H(p,q)+\pi_t\big)
  \right].
  \label{eq:tildeS-bulk}
\end{equation}
Its boundary variation is
\begin{equation}
  \delta \tilde S_{\rm boundary}
  =
  \big[p_i\,\delta q^i - t\,\delta\pi_t\big]_{\sigma_a}^{\sigma_b}.
\end{equation}
Hence, the boundary term vanishes for
\begin{equation}
  \delta q^i(\sigma_a)=\delta q^i(\sigma_b)=0,
  \qquad
  \delta \pi_t(\sigma_a)=\delta \pi_t(\sigma_b)=0,
\end{equation}
while \(\delta t\) is left free at the endpoints.

Stationarity of \(\tilde S\) yields
\begin{align}
  \frac{dq^i}{d\sigma} &= \tilde N(\sigma)\,\frac{\partial H}{\partial p_i},
  &
  \frac{dp_i}{d\sigma} &= -\,\tilde N(\sigma)\,\frac{\partial H}{\partial q^i},
  \label{eq:MJ-eom-qp-from-tildeS}
  \\
  \frac{d\pi_t}{d\sigma} &=0,
  &
  H(p,q)+\pi_t &=0.
  \label{eq:MJ-eom-pit-from-tildeS}
\end{align}
Since \(\pi_t\) is constant along the trajectory, the fixed endpoint values imply
\begin{equation}
  \pi_t=-E,
\end{equation}
for some constant \(E\), the constraint becomes
\begin{equation}
  H(p,q)=E.
  \label{eq:energy-shell-again}
\end{equation}
Thus, the motion is confined to the energy shell, while the time variable drops out of the variational problem.

On shell, \(d\pi_t/d\sigma=0\), so the term involving \(t\,d\pi_t/d\sigma\) vanishes, and the action reduces to:
\begin{equation}
  S_{\rm MJ}[p,q,N;E]
  =
  \int_{\sigma_a}^{\sigma_b} d\sigma\,
  \left[
    p_i(\sigma)\,\frac{dq^i}{d\sigma}
    -N(\sigma)\big(H(p,q)-E\big)
  \right],
  \label{eq:S-MJ-from-root-clean}
\end{equation}
where we have simply renamed \(\tilde N\equiv N\). This is precisely the Maupertuis-Jacobi action in canonical form. Its variation gives:
\begin{equation}
\begin{split}
      \frac{dq^i}{d\sigma}&=N(\sigma)\,\frac{\partial H}{\partial p_i},
  \\
  \frac{dp_i}{d\sigma}&=-\,N(\sigma)\,\frac{\partial H}{\partial q^i},
  \\
   H(p,q)&=E,
\end{split}
  \label{eq:MJ-eom-clean}
\end{equation}
namely Hamilton's equations up to an arbitrary reparametrization, together with the fixed-energy constraint.

The microcanonical description is therefore not based on a fundamentally different dynamics. It corresponds instead to a different boundary-value problem for the same parametrized constrained theory: the endpoint configurations and the conserved energy are fixed, while time is treated as a derived quantity. At the statistical level, this is the classical setting underlying the microcanonical density of states.

The classical lesson is thus already clear. Fixed-time and fixed-energy descriptions need not be viewed as distinct dynamical principles, but as two realizations of the same parametrized constrained system. The quantum construction developed below is the operator and projector counterpart of this classical unity.

\section{Parametrized dynamics and the constrained viewpoint}

The constrained formulation has a standard classical antecedent. Hamiltonian dynamics can be written on an extended phase space in which time is promoted to a canonical variable with conjugate momentum $P_T$. The classical constraint then takes the form:
\begin{equation}
    C=P_T+H(q,p)=0,
\end{equation}
which expresses the arbitrariness of the evolution parameter along physical trajectories. From this viewpoint, fixed-time and fixed-energy descriptions are not two different dynamics, but two different realizations of the same constrained structure.

This classical remark is important only because it clarifies the quantum construction. At the quantum level, the same logic suggests that one should not begin by postulating the canonical and microcanonical ensembles independently at the level of $\hat H$. One should instead identify the common constrained object from which both are recovered.

\section{Extended Hilbert space and constraint projector}

Let $\hat H$ be a self-adjoint Hamiltonian acting on the physical Hilbert space $\mathcal H_{QM}$. We introduce an auxiliary clock Hilbert space $\mathcal H_T$ endowed with operators $\hat T$ and $\hat P_T$ satisfying:
\begin{equation}
    [\hat T,\hat P_T]=i,
    \qquad
    [\hat P_T,\hat H]=0.
\end{equation}
The extended Hilbert space is:
\begin{equation}
    \mathcal H_{\mathrm{ext}}=\mathcal H_T\otimes \mathcal H_{QM}.
\end{equation}
The dynamics is encoded in the single first-class constraint:
\begin{equation}
    \hat C = \hat P_T + \hat H.
\end{equation}
Physical states satisfy:
\begin{equation}
    \hat C\,|\Psi_{\mathrm{phys}}\rangle =0,
\end{equation}
and physical transition amplitudes are obtained by projecting onto the kernel of $\hat C$. At operatorial level, this is implemented by the Dirac delta of the constraint,
\begin{equation}
    \delta(\hat C)=\int_{-\infty}^{+\infty} d\alpha\; e^{i\alpha \hat C}
    =\int_{-\infty}^{+\infty} d\alpha\; e^{i\alpha(\hat P_T+\hat H)}.
\label{eq:deltaC}
\end{equation}
The operator $\delta(\hat C)$ is the fundamental object of the construction. Neither $e^{-\beta\hat H}$ nor $\delta(\hat H-E)$ is taken as primitive. Both are recovered from Eq.~\eqref{eq:deltaC} after choosing a representation in the auxiliary clock sector.

Before proceeding, we should make a remark on the auxiliary clock sector. A potential source of confusion concerns the status of the auxiliary clock operator \(\hat T\). In the present formulation, \(\hat T\) is not introduced to play the role of an operator canonically conjugate to the physical Hamiltonian \(\hat H\) on the system Hilbert space \(\mathcal H_{\mathrm{QM}}\). Rather, \(\hat T\) acts on an independent clock Hilbert space \(\mathcal H_T\), is conjugate only to its own momentum \(\hat P_T\), and commutes with \(\hat H\). The relevant dynamical object is therefore the constraint operator:
\begin{equation}
  \hat{\mathscr C}=\hat P_T+\hat H,
\end{equation}
together with the corresponding projector \(\delta(\hat{\mathscr C})\), rather than a commutator of the form \([\hat T,\hat H]=i\). This is the reason why the present construction is not obstructed by Pauli-type arguments against self-adjoint time operators conjugate to semibounded Hamiltonians. A more detailed discussion is given in Appendix~\ref{app:clock_Pauli}.

\section{Canonical ensemble as clock-time projection}

With the status of the auxiliary clock clarified, we can now state the central point of the construction more explicitly. The constrained projector \(\delta(\hat{\mathscr C})\) provides a single reparametrization-invariant kernel, from which different statistical ensembles are obtained by resolving the clock sector in different representations. In this sense, we show that the canonical and microcanonical ensembles arise as complementary projections of the same constrained quantum dynamics.

We first work in the basis of clock eigenstates $|T\rangle$ and configuration states $|q\rangle$, and define:
\begin{equation}
    |T,q\rangle = |T\rangle\otimes |q\rangle.
\end{equation}
The physical kernel is
\begin{equation}
    K_{\mathrm{phys}}(T_f,q_f;T_i,q_i)=\langle T_f,q_f|\delta(\hat C)|T_i,q_i\rangle.
\end{equation}
Using Eq.~\eqref{eq:deltaC},
\begin{equation}
    K_{\mathrm{phys}}(T_f,q_f;T_i,q_i) =\int d\alpha\;\langle T_f|e^{i\alpha \hat P_T}|T_i\rangle\langle q_f|e^{i\alpha \hat H}|q_i\rangle.
\end{equation}
Since
\begin{equation}
    \langle T_f|e^{i\alpha \hat P_T}|T_i\rangle = \delta(T_f-T_i+\alpha),
\end{equation}
the $\alpha$-integration collapses, and one obtains:
\begin{equation}
    K_{\mathrm{phys}}(T_f,q_f;T_i,q_i)=\langle q_f|e^{i(T_i-T_f)\hat H}|q_i\rangle.
\label{eq:Kphys-time}
\end{equation}
This equation shows that the usual evolution operator appears as a matrix element of the reparametrization-invariant kernel once one chooses the clock-time representation.

The canonical ensemble is recovered by taking a purely imaginary separation in the clock variable,
\begin{equation}
    T_f-T_i=-i\beta.
\end{equation}
Equation~\eqref{eq:Kphys-time} then becomes the Euclidean kernel:
\begin{equation}
    K_\beta(q_f,q_i):=K_{\mathrm{phys}}(T_i-i\beta,q_f;T_i,q_i) =\langle q_f|e^{-\beta \hat H}|q_i\rangle.
\end{equation}
If one identifies the initial and final configurations and integrates over them, one obtains:
\begin{equation}
    Z(\beta)\propto \int Dq\,K_\beta(q,q)=\mathrm{Tr}_{\mathcal H_{QM}}\,e^{-\beta \hat H}.
\end{equation}
Thus, the canonical ensemble is not postulated but recovered from the constrained kernel after three steps: choosing the clock-time basis, imposing an imaginary separation in clock time, and summing over coincident configurations. Temperature is therefore associated with an imaginary lapse of the auxiliary clock variable. In this sense, the canonical ensemble is a particular representational realization of the constrained dynamics rather than its primitive starting point.

\section{Microcanonical ensemble as clock-energy projection}

We now choose the conjugate basis of eigenstates of the clock momentum $\hat P_T$. Let $|E\rangle$ be defined by:
\begin{equation}
    \hat P_T|E\rangle = -E|E\rangle,
    \qquad
    \langle E_f|E_i\rangle = \delta(E_f-E_i).
\end{equation}
The sign convention is chosen so that the constraint $\hat C=\hat P_T+\hat H$ enforces equality between the clock ``energy'' $-P_T$ and the physical energy.

The extended basis states are:
\begin{equation}
    |E,q\rangle = |E\rangle\otimes |q\rangle,
\end{equation}
and the corresponding kernel reads:
\begin{equation}
    K(E_f,q_f;E_i,q_i)=\langle E_f,q_f|\delta(\hat C)|E_i,q_i\rangle.
\end{equation}
Using again Eq.~\eqref{eq:deltaC},
\begin{equation}
    K(E_f,q_f;E_i,q_i)=\int d\alpha\; \langle E_f|e^{i\alpha \hat P_T}|E_i\rangle \langle q_f|e^{i\alpha \hat H}|q_i\rangle.
\end{equation}
Since
\begin{equation}
    \langle E_f|e^{i\alpha \hat P_T}|E_i\rangle=e^{-i\alpha E_i}\delta(E_f-E_i),
\end{equation}
one obtains:
\begin{equation}    
    K(E_f,q_f;E_i,q_i)=\delta(E_f-E_i)
\int d\alpha\;e^{-i\alpha E_i}\langle q_f|e^{i\alpha \hat H}|q_i\rangle.
\end{equation}
The remaining integral is precisely the spectral delta function of the Hamiltonian:
\begin{equation}
    \delta(\hat H-E_i)=\int d\alpha\; e^{i\alpha(\hat H-E_i)},
\end{equation}
so that the kernel becomes:
\begin{equation}
    K(E_f,q_f;E_i,q_i)=\delta(E_f-E_i)\langle q_f|\delta(\hat H-E_i)|q_i\rangle.
\label{eq:microkernel}
\end{equation}
Equation~\eqref{eq:microkernel} shows that the same constrained projector $\delta(\hat C)$, when represented in the conjugate clock-energy basis, reduces directly to the spectral shell operator of the Hamiltonian. The microcanonical density of states then follows as:
\begin{equation}
    \Omega(E)\propto \int Dq\,K(E,q)=\mathrm{Tr}_{\mathcal H_{QM}}\,\delta(\hat H-E).
\end{equation}
From this viewpoint, the microcanonical ensemble is not an auxiliary alternative appended to the theory from outside. It is the complementary projection of the same constrained quantum dynamics in a different clock representation.

\section{Equal footing of the two ensembles}

The main consequence of the construction can now be stated clearly. In the standard formulation, the canonical and microcanonical ensembles appear as formally distinct objects built from $e^{-\beta \hat H}$ and $\delta(\hat H-E)$, respectively. In the constrained formulation, this distinction is secondary. Both objects arise from the same reparametrization-invariant projector $\delta(\hat C)$ and differ only by the representation chosen in the clock sector.

More explicitly,
\begin{itemize}[leftmargin=2em]
\item in the clock-time basis $|T\rangle$, together with an imaginary clock separation $\delta(\hat C)$, it yields the Euclidean kernel and hence the canonical partition function;
\item in the conjugate clock-energy basis $|E\rangle$, the same $\delta(\hat C)$ yields the spectral projector $\delta(\hat H-E)$ and hence the microcanonical density of states.
\end{itemize}
Therefore, canonical and microcanonical statistics are on equal formal footing. What differs between them is not the underlying dynamics, but the way in which the same constrained kernel is projected.

This is the precise sense in which the apparent primacy of the canonical ensemble is representational rather than structural. The canonical ensemble is especially familiar because it is naturally connected to Euclidean time evolution. But within the constrained framework, it is not privileged at the fundamental level. It is one projection among others of the same operatorial structure.

\section{Conclusions}

We have shown that the canonical and microcanonical ensembles need not be introduced as separate statistical constructions. In a parametrized quantum theory with an extended Hilbert space:
\begin{equation}
    \mathcal H_{\mathrm{ext}}=\mathcal H_T\otimes\mathcal H_{QM},
\end{equation}
and constraint:
\begin{equation}
    \hat C=\hat P_T+\hat H,
\end{equation}
both ensembles are recovered as projections of the single reparametrization-invariant kernel $\delta(\hat C)$. In the clock-time representation, one obtains the Euclidean kernel and the canonical ensemble, whereas in the clock-energy representation, one obtains the spectral projector $\delta(\hat H-E)$ and the microcanonical ensemble.

The claim is that the usual asymmetry between canonical and microcanonical ensembles is not built into the underlying dynamics itself. It originates instead from a preferred representational choice. Once the theory is formulated in constrained form, the two ensembles appear as complementary realizations of the same kernel.

This viewpoint is useful precisely because it isolates a simple and robust message: canonical and microcanonical statistics are not independent postulates at the level of $\hat H$, but two projections of one constrained quantum dynamics.



\appendix

\section{Auxiliary clock variable, constrained dynamics, and Pauli's theorem}
\label{app:clock_Pauli}

In this appendix, we clarify the status of the auxiliary clock variable used in the parametrized formulation of the main text and explain why the construction does not conflict with Pauli-type no-go statements on time operators. The key point is that the clock sector is introduced as an independent auxiliary degree of freedom in an extended Hilbert space and that the relevant dynamical object is the constraint operator
\(
    \hat{\mathscr C}=\hat P_T+\hat H
\),
rather than a commutation relation of the form
\(
    [\hat T,\hat H]=i
\).

\subsection{Extended Hilbert space and constrained dynamics}

Let \(\mathcal H_{\mathrm{QM}}\) denote the Hilbert space of the quantum system of interest, with Hamiltonian \(\hat H\) acting on \(\mathcal H_{\mathrm{QM}}\). In the parametrized formulation adopted in the main text, one introduces an auxiliary clock sector \(\mathcal H_T\), carrying a canonical pair \((\hat T,\hat P_T)\), and considers the extended Hilbert space
\begin{equation}
    \mathcal H_{\mathrm{ext}}
    =
    \mathcal H_T \otimes \mathcal H_{\mathrm{QM}}.
\end{equation}
The clock sector is represented in the standard way, with:
\begin{equation}
    [\hat T,\hat P_T]=i,
\end{equation}
while the physical Hamiltonian acts trivially on \(\mathcal H_T\), so that:
\begin{equation}
     [\hat T,\hat H]=[\hat P_T,\hat H]=0.
\end{equation}
Thus, \(\hat T\) is canonically conjugate only to its own momentum \(\hat P_T\), and not to the physical Hamiltonian \(\hat H\).

The dynamics of the extended system is encoded by the first-class constraint:
\begin{equation}
  \hat{\mathscr C}=\hat P_T+\hat H,
  \label{eq:C_constraint_appendix}
\end{equation}
which is the quantum counterpart of the classical reparametrization constraint on the extended phase space. Physical states are generalized solutions of:
\begin{equation}
  \hat{\mathscr C}\,|\Psi_{\mathrm{phys}}\rangle =0,
\end{equation}
and physical amplitudes are defined through the corresponding projector:
\begin{equation}
  \delta(\hat{\mathscr C})
  =
  \frac{1}{2\pi}
  \int_{-\infty}^{+\infty} d\alpha\;
  e^{i\alpha \hat{\mathscr C}}
  =
  \frac{1}{2\pi}
  \int_{-\infty}^{+\infty} d\alpha\;
  e^{i\alpha(\hat P_T+\hat H)}.
  \label{eq:deltaC_appendix}
\end{equation}
This operator implements the constraint in a reparametrization-invariant way and correlates the clock and physical sectors without privileging any particular choice of evolution parameter.

\subsection{Why Pauli's theorem does not apply}

A possible source of confusion is the relation between the auxiliary clock variable \(\hat T\) introduced above and Pauli-type arguments against the existence of self-adjoint time operators. In its standard form, Pauli's observation states that if \(\hat H\) is a semibounded self-adjoint Hamiltonian acting on a Hilbert space \(\mathcal H\), then one cannot, in general, define a self-adjoint operator \(\hat T\) on the same Hilbert space satisfying:
\begin{equation}
  [\hat T,\hat H]=i,
  \label{eq:Pauli_comm_appendix}
\end{equation}
with the usual domain and covariance properties expected of a canonical pair. In particular, the external Schr\"odinger time parameter cannot simply be promoted to an operator canonically conjugate to the physical Hamiltonian on the same physical Hilbert space when \(\hat H\) has a lower-bounded spectrum.

The present construction does not introduce such an operator. Indeed, the
relevant commutation relations are:
\begin{equation}
    [\hat T,\hat P_T]=i,
    \qquad
    [\hat T,\hat H]=0.
\end{equation}
The operator \(\hat T\) acts on the auxiliary factor \(\mathcal H_T\), whereas \(\hat H\) acts on \(\mathcal H_{\mathrm{QM}}\). The physical content of the theory is not encoded in a canonical pair \((\hat T,\hat H)\), but in the constraint:
\begin{equation}
    \hat{\mathscr C}=\hat P_T+\hat H,
\end{equation}
together with the projector \(\delta(\hat{\mathscr C})\) onto its kernel.
Accordingly, the present framework is not obstructed by Pauli's theorem: the latter concerns time operators conjugate to the physical Hamiltonian on the same Hilbert space, whereas here the clock variable is conjugate only to its own auxiliary momentum on an independent factor of the extended Hilbert space.

The role of the clock variable is therefore purely structural. It provides a convenient way of encoding reparametrization invariance and organizing the physical amplitudes in different representations. Its meaning is exhausted by its correlations with the physical degrees of freedom, as captured by matrix elements of the constrained projector, such as:
\begin{equation}
    \langle T_f,q_f|\,\delta(\hat{\mathscr C})\,|T_i,q_i\rangle
\end{equation}
or, equivalently, by mixed representations in which the clock sector is
resolved in the eigenbasis of \(\hat P_T\).

\subsection{Relation to the ensemble projections in the main text}

Once the status of the clock sector is understood in these terms, the
construction used in the main text becomes conceptually transparent. The
canonical and microcanonical ensembles are not introduced through two
independent statistical postulates, but arise as different representations of the same constrained projector \(\delta(\hat{\mathscr C})\) in the clock sector.
In the clock-time representation, one recovers the Euclidean kernel and hence the canonical partition function, while in the conjugate clock-energy representation, one recovers the spectral projector \(\delta(\hat H-E)\) and hence the microcanonical ensemble. The purpose of the present appendix is only to clarify that this mechanism does not rely on a time operator canonically conjugate to the physical Hamiltonian, and is therefore fully compatible with Pauli-type no-go statements.

\end{document}